\documentstyle[12pt,aaspp4,flushrt]{article}

\newcommand{\msb}{$M_{\odot}$~}
\newcommand{\ms}{$M_{\odot}$}

\newcommand{\ct}{$^{13}$C}

\newcommand{\ctb}{$^{13}$C~}

\newcommand{\neon}{$^{22}$Ne}

\begin{document}

\title{Lead: Asymptotic Giant Branch production and Galactic Chemical
Evolution}

\author{Claudia Travaglio\altaffilmark{1}}
\affil{1. Max-Planck Institut f\"ur Astronomie, K\"onigsthul 17,
D-69117 Heidelberg, Germany, claudia@mpia-hd.mpg.de}
 
\author{Roberto Gallino\altaffilmark{2}}
\affil{2. Dipartimento di Fisica Generale, Universit\`a di Torino, Via
P.Giuria 1, I-10125 Torino, Italy, gallino@ph.unito.it}
 
\author{Maurizio Busso\altaffilmark{3}}
\affil{3. Osservatorio Astronomico di Torino, Strada Osservatorio 20,
I-10025 Torino, Italy, busso@to.astro.it}
 
\author{Raffaele Gratton\altaffilmark{4}}
\affil{4. Osservatorio Astronomico di Padova, Vicolo dell'Osservatorio 5,
I-35122 Padova, Italy, gratton@pd.astro.it}

\begin{abstract}
The enrichment of Pb in the Galaxy is followed 
in the framework of a detailed model of Galactic chemical evolution
that already proved adequate to reproduce the chemical enrichment of O and
of the elements from Ba to Eu. The stellar yields are computed through
nucleosynthesis calculations in the Asymptotic Giant Branch (AGB) phase
of low- and intermediate-mass stars, covering a wide range of
metallicities. The physical parameters of the stellar structure were 
derived from full stellar evolutionary models previously computed. We show that
low-mass AGB stars are the main producers of Pb in the Galaxy, with a 
complex dependence on  metallicity and a maximum efficiency  at
[Fe/H] $\sim$ $-$1. Our calculations succeed in reproducing the abundances of
Pb isotopes in the solar system:
the role attributed by the classical analysis of the $s$ process to the
{\it strong component}, in order to explain more than 50\% of solar $^{208}$Pb,
is actually played by the high production of Pb in low-mass
and low-metallicity AGB stars. We then follow the
Galactic chemical evolution of Pb isotopes and give our
expectations on the $s$-process contribution to each of them at the epoch of
the solar system formation. Finally, we present new spectroscopic estimates 
of Pb abundance on a sample of field stars and compare them, together
with a few other determinations available, with the
predicted trend of [Pb/Fe] in the Galaxy.

\end{abstract}


\section{Introduction}

Neutron captures are the main way 
for the building up of the nuclei heavier than iron. According to the seminal 
work by Burbidge et al.~(1957), in order to reconstruct the
Galactic evolutionary history of heavy elements one has to consider
two major mechanisms of neutron addition: the $s$-process (slow
neutron-captures) and the $r$-process (rapid neutron-captures). For the
$s$-process, the abundance distribution in the solar system was early
recognized to derive from a non unique site. Indeed, the phenomenological 
approach of $s$-process studies, or {\it classical} analysis, 
was shown to require three components to account for the
abundances of all $s$-nuclei from Fe to Bi: the {\it weak}, the {\it
main}, and the {\it strong} component (Seeger, Fowler,
\& Clayton~1965; Clayton \& Rassbach~1967; K\"appeler et al.~1982;
K\"appeler, Beer, \& Wisshak~1989).

The weak component, responsible for
the $s$-process nuclides up to $A \simeq$ 90, was ascribed 
to neutron captures occurring in advanced evolutionary phases of massive stars, 
through the activation of the $^{22}$Ne($\alpha$,n)$^{25}$Mg reaction (Lamb 
et al.~1977; Prantzos et al.~1990; Raiteri et al.~1993 and references therein).
The main component, feeding the heavier $s$-process nuclides up to $A =$ 
208, is known to originate in AGB stars of Galactic disk metallicity
during recurrent thermal pulses, hereafter TP. Many theoretical and
observational works (see Busso, Gallino, \& Wasserburg~1999 for a review)
converge on the idea that Low-Mass Stars (hereafter LMS, $M$ $\le$ 4 \ms)
play the dominant role in $s$-processing, and that  neutrons are released
in radiative conditions in the interpulse phase via the
$^{13}$C($\alpha$,n)$^{16}$O reaction (Straniero et al.~1995). A tiny
$^{13}$C {\it pocket} is thought to develop in the top layers of the He- and
C-rich zone (called He intershell) as a consequence of the penetration of a
small amount of protons from the envelope, to burn locally and to produce
neutrons before the  development of the next TP. During advanced TPs a second 
small neutron burst is released by the marginal activation of the
$^{22}$Ne($\alpha$,n)$^{25}$Mg reaction. 
Finally, the strong component was advanced on theoretical grounds by the
classical analysis of the $s$-process (Clayton \& Rassbach 1967), 
in order to reproduce more than 50\% of $^{208}$Pb, 
the most abundant Pb isotope in the solar system,
although the astrophysical site for this process remained unknown.

In this work we study the Galactic enrichment of Pb in the light of
AGB nucleosynthesis results presented by Straniero et al.~(1997)
and Gallino et al.~(1998). One of the main goals of these works
was the development of a new set of model calculations for AGB stars with
different masses and metallicities, to provide stronger evolutionary supports 
to nucleosynthesis calculations. The results demonstrated in detail
the dependence of the $s$-process yields on stellar metallicity, and 
introduced the idea that a separate astrophysical site for the strong
component might be unnecessary (Gallino et al.~1998), because a large 
production of $^{208}$Pb derives from AGB stars of low metallicity. 
This expectation must however be substantiated with a  detailed model for
the Galactic enrichment, to verify whether the production of Pb isotopes
is really adequate. 
 
The paper is organized as follows: in \S~2 we discuss the sensitivity
of Pb production (in particular $^{208}$Pb) to metallicity and to 
different assumptions on the $^{13}$C pocket in low-mass AGB stars. We
also analyze the role of intermediate-mass AGB stars in the Pb production
at different metallicities. In \S~3 we present spectroscopic
estimates by synthetic spectra of the Pb abundance on a limited sample of
field metal-poor stars using the Pb line at 3683.48~\AA~ and outline the
difficulties and the aims of such an analysis.
In \S~4 we briefly introduce the Galactic Chemical Evolution (hereafter
GCE) model adopted, present our predictions for
the $s$-components of the Pb isotopes in the solar system, and compute
the [Pb/Fe]\footnote{We adopt the usual spectroscopic notation [Pb/Fe] =
log (Pb/Fe)
$-$ log (Pb/Fe)$_{\odot}$} trend vs. metallicity in the three Galactic
zones in which the GCE model is organized (halo, thick disk and thin disk). 
Then we compare our model predictions with the
above new spectroscopic determinations and with a few other published
data, outlining the need of a dedicate observational effort. Finally, in
\S~5 we summarize the main conclusions and point out a few aspects
deserving further analysis.

\section{Pb production by $s$-process nucleosynthesis}

We compute stellar yields for $s$-processing of Pb in AGB stars as a
result of post-process calculations on the basis of stellar evolutionary models
obtained with FRANEC (Frascati Raphson-Newton Evolutionary Code; see
Chieffi \& Straniero~1989) and presented in Straniero et al.~(1997) and 
Gallino et al.~(1998). Additional models have been discussed by Vaglio et
al.~(1999) and Straniero et al.~(2000). 
The detailed  evolutionary computations were performed  at [Fe/H] = 0.0,
$-$0.3 and $-$1.3 for the AGB stars of mass 1.5 and 3 \ms. For the 5 \msb
case the metallicities were [Fe/H] = 0.0 and $-$1.3, while for the 7 \msb
star only models of solar metallicity were calculated. 
The $s$-process results has then been obtained with 
post-process runs span over the metallicity interval from $Z =
Z_\odot$ down to $Z = Z_\odot$/4000 for 1.5, 3, 5 and 7 \ms. The results
of the FRANEC stellar structures are extrapolated to the whole metallicity
range adopted for the $s$-process calculations (more details are available
in Straniero et al.~1997, ~2000, and Gallino et al.~1998).
Mass loss was taken into account by adopting
the parameterization of Reimers~(1975). The values adopted for this work
are $\eta =$ 0.3 for 1.5 \msb, and $\eta =$ 1.5 for 3 \msb (Straniero et
al.~1997). Calculations for 5 \msb and 7 \msb 
are instead performed with $\eta =$ 10 and $\eta =$ 3, respectively 
(Straniero et al.~2000). The models self-consistently reproduce the
so-called {\it third dredge up} episode (TDU), defined as the penetration
of the convective envelope below the H-He discontinuity. TDU was found to
occur after a limited number of TPs and to cease when the envelope mass
decreases by mass loss below a limiting mass of $\sim$ 0.5 \ms. 

The cumulative mass for the He intershell dredged up into the envelope of AGB 
models of 1.5, 3, 5, and 7 \msb according to FRANEC models and eventually
ejected into the interstellar medium by stellar winds is reported in
Table~1. 
Different treatments for convection with respect to the one adopted in
FRANEC, where the convective/radiative borders are followed according to the
Schwarzschild criterion, may provide larger TDU efficiencies, 
as e.g. discussed  by Frost \& Lattanzio (1996),
Herwig et al.~1997; Langer et
al.~1999; Herwig~2000). Moreover, the criterion adopted for 
the mass loss during the AGB phase introduces a second major source of 
uncertainty, and at the present moment only the comparison with
observations is expected to provide constraints on these points.

\begin{table}[t]
\begin{center} TABLE 1\\
{\sc Cumulative mass (in~\ms) \\
dredged-up from the He intershell \\
to the envelope of AGB stars}\\
\vspace{1.0em}
\begin{tabular}{rllll}
\hline\hline
 [Fe/H]  & 1.5 \msb  & 3 \msb & 5 \msb & 7 \msb \\
\hline
$-$1.3 & 0.07 & 0.16 & 0.20 & 0.20 \\
$-$0.5 & 0.04 & 0.09 & 0.20 & 0.20 \\ 
 0.0 & 0.02 & 0.05 & 0.10 & 0.10 \\
\hline
\hline
\end{tabular}
\end{center}
\end{table}

Despite first successful models for the formation of the \ctb pocket
have already been presented (Hollowell \& Iben 1988; 
Herwig et al.~1997; Langer et
al.~1999), the mass involved and the profile of $^{13}$C formed at
TDU in the He intershell are still to be assumed as free parameters, being
related to hydrodynamical mixing mechanisms that cannot be properly
accounted for in stellar models (Straniero et al.~2000). However a series of
constraints 
can be obtained by comparing spectroscopical abundances in Pop.~I AGB stars and in
Pop.~II  Ba and CH stars  (e.g., Smith \& Lambert~1990; Luck \& Bond~1991;
Vanture~1992; Plez, Smith, \& Lambert~1993; van Winckel \& Reyniers~2000)
with model predictions. In fact, Galactic disk AGB stars show a
significant spread in $s$-process efficiencies that can be attributed to variations 
in the $^{13}$C amount in the pocket and to the large sensitivity of the 
resulting $s$-process
distribution on the initial metallicity.
As discussed by Busso et al.~(1995,~2000), the observed data are
distributed around an average $s$-process efficiency rather similar to that of
the main component in the solar system. However, the considerable
intrinsic scatter revealed by spectroscopic studies can only fix broad
limits to the amount of \ctb burnt. 
Notice that the same spread in \ctb concentrations
was found to be appropriate to account for the $s$-process isotopic
signatures of heavy elements in presolar grains recovered  in meteoritic
material (Zinner~1997), most likely condensed in the circumstellar envelopes
of AGB stars of close to solar metallicity (Gallino et al.~1993,~1997; Lugaro
et al.~1999; Amari et al.~2000). In our calculations the intrinsic spread in
the $s$-process yields at each metallicity was modelled parametrically by
varying the  $^{13}$C concentration in the pocket by factors from 0 to
2 times the ``standard'' (ST) value discussed in  Gallino et al.~(1998)
and
Arlandini et al.~(1999) ($\sim$ 4 10$^{-6}$ \msb of $^{13}$C). 
The ST choice was shown to reproduce the main component 
through AGB models in the mass range 
1.5 $-$ 3 \msb and metallicity [Fe/H] = $-$0.3. 
Identical best-fits to the solar component are obtained for other
choices of the \ctb pocket, provided the metallicity varies according
to the rule $N$(\ct)/$N$($^{56}$Fe) = const. Implications for different
choices of the ST profile and/or metallicity have  been discussed
elsewhere (Busso et al.~1995,~2000). 

Following a similar approach, Travaglio et al.~(1999) analyzed
the GCE of Ba, La, Ce, Nd, Pr, Sm, and Eu. 
Both here and in Travaglio et al.~(1999) unweighted yields averaged over the
whole set of \ctb abundances have been adopted. In Fig.~1 the abundances
by mass of $^{208}$Pb in the He intershell material cumulatively mixed with
the surface of a 1.5 \msb star by
TDU episodes are shown as a function of metallicity. The main difference
in the present computations with respect to those presented in Travaglio
et al.~(1999) lies in a finer interpolation grid at the
lowest concentrations of \ct. This larger grid of models can be useful
for a better account of recent observational constraints from evolved
stars at low
and intermediate metallicity (Ryan et al.~2000a; Travaglio, Gallino, 
\& Busso~2000). Actually, the extension of the $^{13}$C
parameterization towards the lowest concentration of $^{13}$C has
negligible influence on the average unweighted yields (see \S~3). 

The strong dependence of the $s$-process yields on  stellar metallicity is
evident in Fig.~1. The trend of $s$-process elements,
and of $^{208}$Pb in particular, the most abundant among Pb isotopes, can be
understood as follows. The build up of heavy $s$-process nuclei is due to
neutron captures starting on pre-existing seeds; it is therefore formally of
secondary origin and one would expect them to decline with
declining metallicity. However, the abundance distribution is not only
dependent on the initial Fe concentration, but also on the neutron addition
efficiency, i.e. on the neutron exposure $\tau$ (where $\tau \equiv \int n_n
v_T dt$, with $n_n$ neutron density and $v_T$ thermal velocity). The gradual
increase of the neutron exposure toward low metallicities
entirely masks this expected secondary behaviour (as first noticed by
Clayton 1988), resulting in a rather complex dependence of
$s$-process yields on metallicity.  Starting from AGB stars of nearly
solar metallicity and going toward more metal-poor stars, the
neutron-flux first builds up the $s$-elements belonging to the Zr-peak
(at neutron magic number $N$ $=$ 50). Then the Zr-peak is partly bypassed
and the production of
the elements at the second $s$-peak increases ($N$ $=$ 82, Ba-peak),
reaching a maximum at  [Fe/H] $\sim$~$-$0.6 (Travaglio et al.~1999; Busso
et al.~1999). For even lower metal contents, the $n$-flux feeds
Pb (in particular $^{208}$Pb), with a maximum production yield at [Fe/H] 
$\simeq -$1 (see Fig.~1). The value of [Fe/H] at which the maximum
is reached depends on the choice of the $^{13}$C-pocket. For lower
$^{13}$C concentrations the production peak is shifted towards
lower metallicities. Finally, when very low metallicities are reached,
also Pb starts to decrease, but it never really follows the expected secondary
behavior. In fact, at very low metallicities,  after the shortage of Fe
during $^{13}$C consumption becomes a conspicuous effect, 
the $s$-elements can continue to be fed thanks to the neutron captures
starting on the abundant lighter primary neutron absorbers (e.g. $^{16}$O,
$^{22}$Ne and its progeny).

We also considered in the GCE model the role of intermediate mass AGB stars. 
In Fig.~2 we show a
plot similar to Fig.~1, but for the 5 \msb ({\it upper panel}) and the 7 \msb
({\it lower panel}) models. In these stars, the $^{22}$Ne($\alpha$,n)$^{25}$Mg
reaction is efficiently activated (Iben~1975; Truran \& Iben~1977),
since the temperature at the base of the convective
pulse reaches values of $T =$  3.5 10$^8$ K. 
As a consequence, the neutron exposure by the 
\neon~ neutron source becomes more significant. Also the
peak neutron density during the TP phase is consistently higher than in
LMS ($\rho\sim10^{11}$ n/cm$^3$, see Vaglio et al.~1999, Straniero et
al.~2000), overfeeding a few neutron-rich isotopes involved in important
branchings along the $s$ process, such as $^{86}$Kr, $^{87}$Rb, and
$^{96}$Zr,
with respect to the production of the $s$-only nuclei. As for the choice
of the \ctb neutron source, due to the much shorter interpulse phases in
these stars ($\sim$6500 yr for a 5 \msb
and $\sim$1500 yr for a 7 \ms) with respect to LMS ($\sim$ 3 $-$ 6 10$^4$ yr), the
He intershell mass involved is smaller by one order of magnitude.
Consequently, also the TDU of $s$-process-rich material from the He
intershell into the surface is reduced, again by roughly one order of
magnitude. Due to the above reasons, for the 5 \msb and 7 \msb cases we
considered as a new ``standard'' choice (STIMS) a $^{13}$C mass 
scaled accordingly ($M$($^{13}$C)$_{\rm STIMS}$ = 10$^{-7}$ \ms). 
Then, in the calculations we considered a spread of \ctb abundances by a 
factor $\pm$2 (from STIMS$\times$2 down to STIMS/2), as shown in Fig.~2. 
On the whole, IMS play a minor role in the
production of Galactic Pb. A more extended set of evolutionary calculations 
are needed for IMS, however the model of 5 \msb and [Fe/H] = $-$1.3 does
not show any relevant differences with respect to the one computed at
solar metallicity (Straniero et al. 2000).

\section{Lead abundances in stars}

Lead has very few lines suitable for an abundance analysis in the
optical range. The only relatively unblended line in the solar spectrum is
the Pb~I line at 3683.48~\AA~ (Grevesse 1969; Hauge \& S{\o}rli 1973;
Youssef \& Khalil 1989). A resolution $R$ $\geq$ 50,000\ is
required to adequately separate it from the nearby Fe~I blend at
3683.623~\AA. Few spectra of high S/N for metal-poor stars have been taken in
this difficult spectral region at this resolution. A line of similar
strength is located at 4057.81~\AA; however this line is more heavily
blended (mainly with CH lines), leading to more uncertain abundances.

Hence, very few abundance determinations exist for Pb.
Sneden et al. (1998) used UV spectra obtained with HST to derive Pb
abundances
in a few halo stars with large excesses of n-capture elements. The Pb~I line
at 2833.05~\AA~ was clearly seen in HD~126238; a tentative detection was
obtained in HD~115444, while the line was not detected in the ``normal''
star
HD~122563. Sneden et al. (2000) measured a Pb abundance from Keck spectra of
CS~22892-52. Finally, Aoki et al.~(2000) and Ryan et al.~(2000a) recently
gave a Pb abundance for the C-rich and $s$-rich star LP~625-44 from the
line at 4057.81~\AA.

\begin{table}[t]
\begin{center} TABLE 2\\
{\sc Pb abundances in 10 metal-poor stars}\\
\vspace{1.0em}
\begin{tabular}{lccc}
\hline
Star       & [Fe/H]  & [Ba/Fe] & [Pb/Fe] \\
\hline
HD~2665    & $-$1.95 & $-$0.32 & $<$1.0  \\
HD~6229    & $-$1.01 & $-$0.04 &    0.3: \\
HD~19445   & $-$1.97 & $-$0.08 &    ..   \\
HD~21581   & $-$1.57 & $-$0.06 & $<$0.5  \\
HD~23439A  & $-$0.99 &   +0.21 &    0.6  \\
HD~175305  & $-$1.31 &   +0.01 &    0.3: \\
HD~194598  & $-$1.05 & $-$0.02 &    ..   \\
HD~201891  & $-$1.04 & $-$0.04 & $<$0.5  \\
BD~+23~3912 & $-$1.40 &   +0.13 &    ..   \\
BD~+29~0366 & $-$0.92 & $-$0.08 &    0.3: \\
\hline
\end{tabular}
\end{center}
\end{table}

To improve this situation, we used a set of spectra obtained in the last
years at McDonald Observatory for another program (Gratton et al. 2000).
These spectra have a resolution of $R$ $\sim$ 60,000. For ten of
the stars considered in that program, the observed spectral range extended
down to $\sim$ 3678~\AA, allowing observation of the best Pb~I line at
3683.48~\AA. These spectra are shown in Fig.~3, where they are compared
with synthetic spectra obtained with the atmospheric parameters listed by
Gratton et al. (2000), and three different abundances of Pb: [Pb/Fe] = 0.0,
0.5, and 1.0. A mark signs the location of the Pb~I line. In our spectra
this is clearly out of the wings of the  Fe~I blend at 3683.623~\AA.
Table~2 lists the abundances we inferred from these spectra; additionally,
we also give abundances for Ba, derived from the same spectra. In most
cases Pb abundances
are uncertain, since the noise level is quite high at these wavelengths that
are at the UV extreme of the spectra. Pb is clearly detected in the spectrum
of HD~23439A, a star possibly classified as a CH star (Tomkin \& Lambert
1999), with small excesses of C and Ba ([C/Fe] = +0.09, [Ba/Fe] = +0.21).
For this star we derive a quite large excess of Pb: [Pb/Fe] = +0.6. We think
the Pb line is also present in the spectra of HD~6229, HD~175305, and
BD~+29~366, although in all cases detection is uncertain. We did not
detect the Pb line in the spectra of the remaining stars; upper limits
were obtained for HD~2665, HD~21581, and HD~201891.

Notice that according to Tomkin \& Lambert (1999) HD~23439A belongs to a
binary system of which  HD~23439B (a single-lined spectroscopic binary) is
also a dwarf CH star with a quite similar $s$-element composition.
According to the authors
the $s$-element enhancements must be primordial, reflecting a heterogeneous
local interstellar medium. Unfortunately this companion does not belong to
the sample of stars by Gratton et al.~(2000). Analysis of the Pb~I line in a
spectrum in this star of comparable quality is highly desirable.

\section{Galactic chemical evolution of Pb} 
 
The GCE model adopted here is described in detail in Ferrini et al.~(1992) and
Travaglio et al.~(1999). As mentioned, it is based on the interconnected
evolution of three zones: halo, thick disk and thin disk, whose relative
composition in stars, gas phases, and stellar remnants is followed during
the Galactic age. Here we consider the evolution of the solar annulus,
located at 8.5 kpc from the Galactic center. The Star Formation Rate (SFR)
is the outcome of self-regulating processes occurring in the molecular gas
phase, either spontaneous or stimulated by the presence of massive stars.
In Fig.~4 (small box) the SFR is plotted as a function of [Fe/H]
during the evolution of the Galaxy (see also Travaglio et al.~1999).

When introducing the $s$-process stellar yields of AGB stars only, 
we get the results listed in Table~3 (sixth column)
for the Galactic $s$-process contribution to the various Pb isotopes at
the epoch of  solar system formation. In particular, at $t = t_\odot$ we
obtain an $s$-fraction to Pb of 91\%. The uncertainty in this evaluation
may depend on the set of prescriptions adopted to estimate the
$s$ yields from AGB stars with varying the metallicity and on the general
prescriptions adopted in the GCE model. However, as recalled before, a
high degree of consistency is inferred from the detailed comparison of
the predicted yields for the elements from Ba to Eu with
spectroscopic determinations in unevolved stars in the Galaxy (Travaglio
et al.~1999). Concerning Pb itself, its $s$-fraction to solar depends on
the present estimate of solar Pb based on meteoritic studies, which,
according to Anders \& Grevesse, is given at 7.8\% (one s.d.). In turn,
the effect of the uncertainty of neutron capture cross sections on Pb
isotopes is very small, because of the high precision of those
determinations (see Bao et al.~2000).  

In the second column we also report, as reference, the elemental solar
abundance of lead from Anders
\& Grevesse~(1989). In the third column we show the
prediction by Arlandini et al.~(1999), as derived from the best-fit to the
main component of the $s$-process in the solar system with the stellar
model. The best-fit reported 
in this column was obtained with AGB models of metallicity [Fe/H] =
$-$0.3 and the ST choice, by averaging the production in the mass
range between 1.5 and 3 \ms. In the fourth column the updated predictions
by the classical analysis are given (from Arlandini et al.~1999).
Finally, in the fifth column we report the prediction from Beer,
Corvi, \& Mutti~(1997) for the strong component of the phenomenological
approach.  

\begin{table}[t]
\begin{center} TABLE 3\\
{\sc $s$-process fractional contributions at} t=t$_\odot$\\ {\sc with
respect to solar system abundances}\\
\vspace{1.0em}
\begin{tabular}{rlllll}
\hline\hline
  & Solar$^{(\rm a)}$ & Main-$s$$^{(\rm b)}$ & Main-$s$$^{(\rm c)}$ &
Strong$^{(\rm d)}$ & GCE$^{(\rm e)}$ \\
  & (\%) & (\%) & (\%) & (\%) \\
\hline
$^{204}$Pb & 1.94  & 94  & 79 & 2 & 90 \\
$^{206}$Pb & 19.20 & 58  & 31 & 2 & 60 \\
$^{207}$Pb & 20.62 & 64  & 30 & 3 & 77 \\
$^{208}$Pb & 58.30 & 34  & 10 & 56 & 89 \\
\hline
Pb & & 46 & 19 & & 91 \\
\hline
\hline
\end{tabular}
\end{center}
\hspace{9em}$^{(\rm a)}$ -- Anders \& Grevesse~(1989)
\vspace{0.1em}

\hspace{9em}$^{(\rm b)}$ -- Arlandini et al.~(1999), ``stellar component''
\vspace{0.1em}

\hspace{9em}$^{(\rm c)}$ -- Arlandini et al.~(1999), ``classical
component''
\vspace{0.1em}

\hspace{9em}$^{(\rm d)}$ -- Beer, Corvi, \& Mutti~(1997)
\vspace{0.1em}

\hspace{9em}$^{(\rm e)}$ -- This work
\end{table}

From a comparison of the best fit by the classical analysis to the main 
component by 
Arlandini et al.~(1999) with the outcome of our GCE, one can notice that
there is a good agreement for all Pb isotopes, with the exception of
$^{208}$Pb. In fact
Arlandini et al.~(1999) obtained an $s$-process contribution of 10\% to
$^{208}$Pb (19\% to elemental Pb). They instead  estimated a higher 34\% of
the $s$-process
contribution to $^{208}$Pb (and 46\% to elemental Pb) 
from their single AGB model (``stellar model'') with the ST choice and
[Fe/H]= $-$0.3 (actually
a mean of 1.5 and 3 \msb AGB models), which also reproduces the main component. 
The discrepancy of the two above predictions on $^{208}$Pb by the classical
analysis and by the ``stellar model'' was already a clear indication that
not a unique AGB stellar model, nor the classical analysis, were able to
explain the main component in the solar system, which must be considered
as the outcome of different generations of AGB stars prior of the solar
system formation.

As a matter of fact, our GCE calculations provide 89\% of $^{208}$Pb abundance, 
thanks to the contribution of different generations of AGB stars and in 
particular to those at low metallicities, which are the main
contributors to $^{208}$Pb. The proper account of GCE 
confirms in a quantitative way what was anticipated
by Gallino et al.~(1998), i.e. that the role previously attributed to the
strong component is actually played by AGB stars of low metallicity.  

A remarkable result of introducing the AGB yields at different
metallicities (as shown in Fig.~1 for $^{208}$Pb)
as input data of GCE computations is that the predicted
chemical compositions in the interstellar medium at any epoch are grossly independent
of the particular $^{13}$C {\it profile} adopted inside the $^{13}$C pocket.
For example, the results presented in the last column of Table~3 remain
essentially the same even assuming a constant $^{13}$C mass fraction in
the pocket. A more detailed analysis will be presented elsewhere,
including the GCE trends of all Pb isotopes.

The $s$-process yields discussed in \S~2 and shown in Fig.~1 
allow us to estimate the chemical enrichment of Pb in the Galaxy
at different epochs. We show in Fig.~4 the
resulting $s$-fraction of lead in the thick and in the thin disk.
This has been obtained by considering AGB stars in the mass range 2 $-$ 8
\ms, but the dominant production of Pb comes in the mass range 2 $-$ 4
\msb(see discussion in \S~2). It is evident that the
$s$-process contribution dominates the GCE of Pb, starting from [Fe/H]
$\simeq -$1.5, both in the thick and thin disk. This is mostly due to the
coincidence between the peak of Pb production in LMS (in the range
$-1.5 \le$ [Fe/H] $\le -$1) with the peak of the star formation rate 
in the thick and thin disk. 
At lower values of [Fe/H], independently of the characteristics of
the GCE model, the contribution of $s$-process
nucleosynthesis rapidly decreases as a result of the long lifetimes of  
AGB stars. 

This confirms previous results of the GCE
evolution of elements from Ba to Eu (Travaglio et al.~1999). At very low
metallicities, as suggested by spectroscopic observations (in particular of Ba
and Eu) and  pointed out by Travaglio et al.~(1999), the
$s$-contribution by AGB stars is by far too small.
Instead, as anticipated by Spite \& Spite~(1978) on observational grounds
and by Truran~(1981) on theoretical grounds,
the heavy element abundance pattern in very metal-poor stars is
compatible with an $r$-process origin. This point was recently sustained by 
observations of low metallicity halo stars (Sneden et al.~1998; McWilliam~1998; 
Sneden et al.~2000; Westin et al.~2000; Ryan et al.~2000b, and 
references therein). 

From the theoretical point of view, despite a large number of recent
works, the
astrophysical site of the $r$-process is still a debated problem (e.g., 
Wheeler, Cowan, \& Hillebrandt~1998; Freiburghaus et al.~1999). In order
to quantify the $r$-contribution, as described in more detail by
Travaglio et al.~(1999), we treat the
$r$-process as a typical primary mechanism occurring in a subset of Type II
SNe (those in the mass range 8 $-$ 10 \ms). Recent theoretical
simulations by Wheeler et al.~(1998) support this choice. Our estimate of
$r$-process abundances at $t = t_\odot$ was simply derived by subtracting 
from the solar abundances the $s$-fractions.

In Fig.~4 the predicted Galactic evolutionary trend for the $s + r$
contribution to Pb is shown. In the figure we also compare our model to recent 
spectroscopic data for halo and disk field stars at different
metallicities, discussed in the previous Section.
It is clear that a larger set of observations
is desirable, and we underline that these are the first
available observational data of Pb. 
The star with [Fe/H] = $-$3.1 and [Pb/Fe] = +1.3 is the very
peculiar  CS~22893-052 (recent observations of Pb in this star has been
presented by Sneden et al.~2000). This star shows exceptionally high
abundances of all heavy elements, with $r$-process/Fe enrichment of
$\sim$40 times the solar ratios.
Therefore it was probably born in an environment strongly polluted by
SN debris with respect to the average interstellar gas in the halo.  This
supports the current idea of an
incomplete mixing of the gas in the Galactic halo at early epochs, which would
allow the formation of stars both very deprived, or very enriched, in
$r$-process elements (for recent studies of this problem see e.g. Tsujimoto,
Shigeyama, \& Yoshii~1999; Raiteri et al.~1999; Argast et al.~2000; Travaglio,
Galli, \& Burkert~2001). 

\section{Conclusions}

In this paper we have studied the evolution of Pb in the interstellar gas
of the Galaxy, adopting a new set of models for $s$-processing in
AGB stars of different metallicities, and new observational constraints 
of unevolved field stars for Pb abundances.

We have calculated the $s$-process yields with post-process calculations
based on AGB models and applied these yields in the framework of a Galactic
model that follows the GCE of the halo, thick and thin disk. Taking into
account the important role played by different generations of
low-metallicity AGB stars for the production of Pb in the Galaxy, we obtain
at $t = t_\odot$ a Pb $s$-fraction  of 91\%, together with a sharp
increase at [Fe/H] $\simeq -$1.5. Confirming the prediction by Gallino et
al.~(1998), we also found that the
production of the most abundant Pb isotope, $^{208}$Pb, is matched when the
$s$-process occurring in low-metallicity AGB stars is
properly considered in the context of the chemical evolution of the Galaxy. 
Only a small portion of $^{208}$Pb can be attributed to the $r$-process,
even accounting for decays of transuranic isotopes.  

{\acknowledgements}
We are indebted to F. Ferrini for allowing us free access to his GCE
model, to O. Straniero, A. Chieffi and M. Limongi for sharing the results
of FRANEC computations, and for a longstanding collaboration on
scrutinizing deeper and deeper in AGB stellar evolutionary models. We also
thanks S.G. Ryan for very useful discussions.

\newpage

\plotone{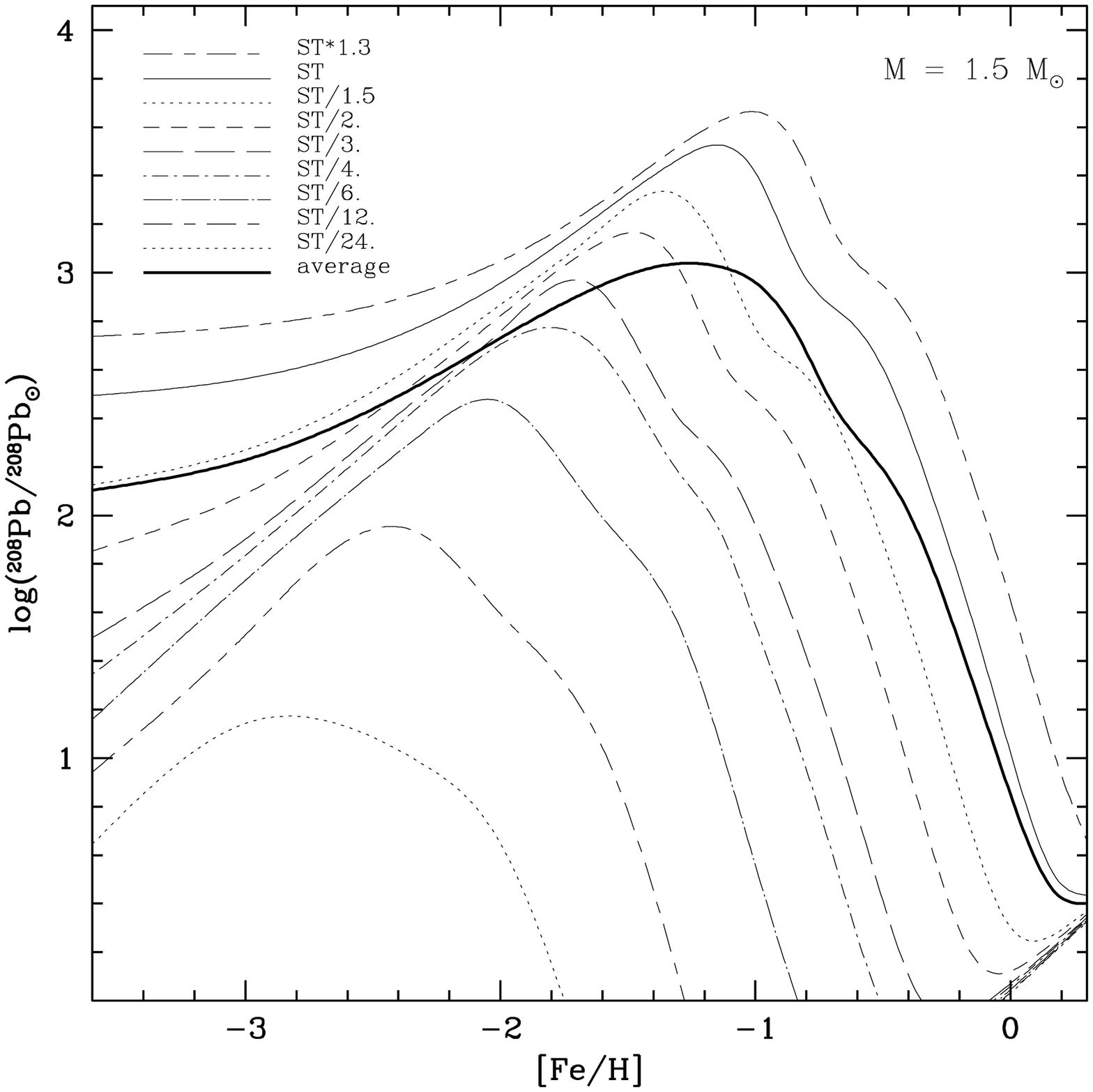}

\figcaption[fig1.eps]{Abundances by mass (relative to solar) of $^{208}$Pb 
in the He intershell material cumulatively mixed with the surface of a
1.5~\msb star by
third dredge up episodes as function of metallicity, for different
assumptions on the mass of the $^{13}$C pocket. The {\em thick continuous 
line} represents the unweighted average of all cases shown.\label{fig1}}

\newpage

\plotone{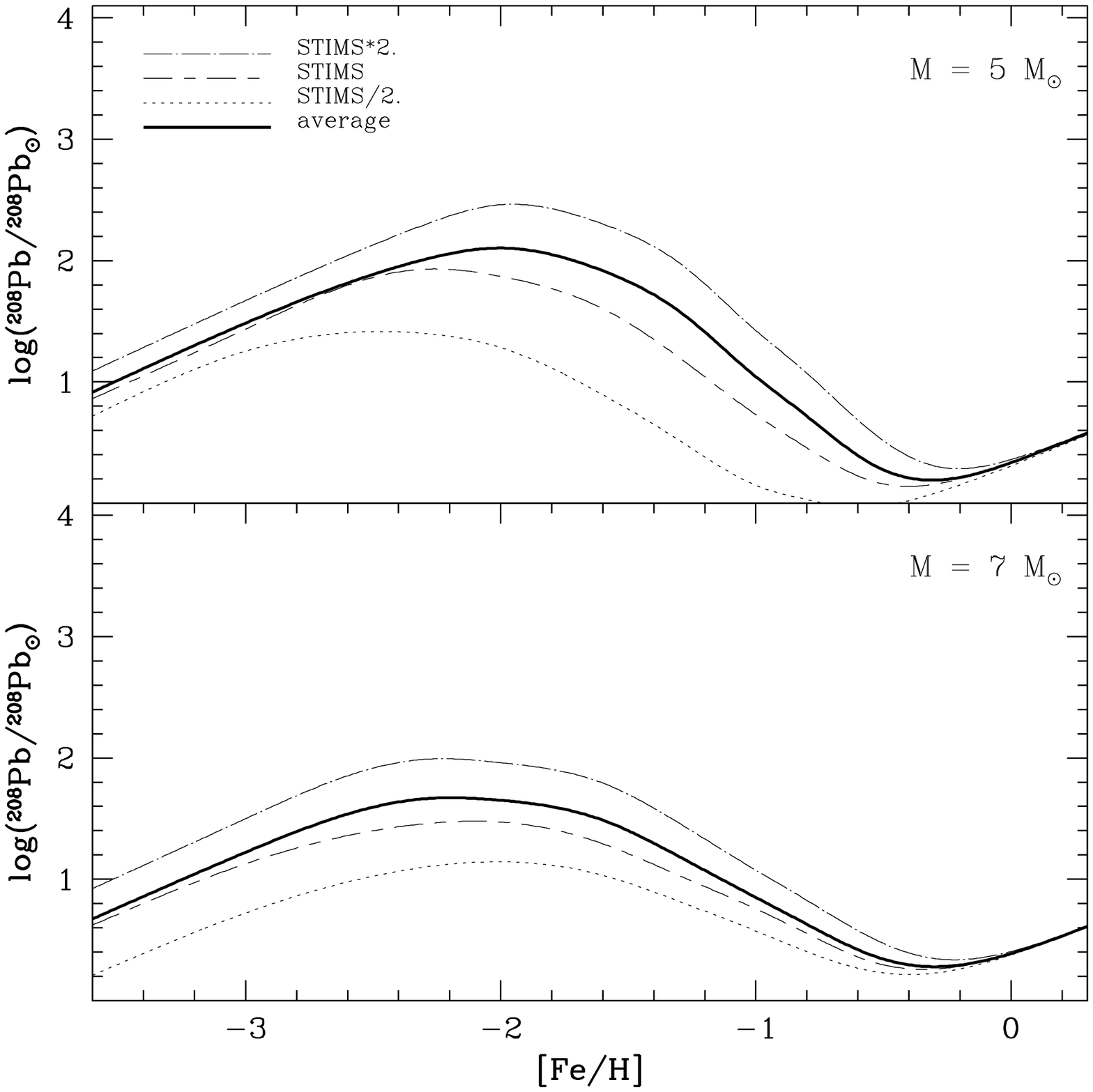}

\figcaption[fig2.eps]{The same of Fig.~1, but for a 5 \msb star ({\it 
upper panel}), and for a 7 \msb star ({\it lower panel}). In both
panels the {\em thick continuous lines} represent the unweighted average
of the cases shown.\label{fig2}}

\newpage

\plotone{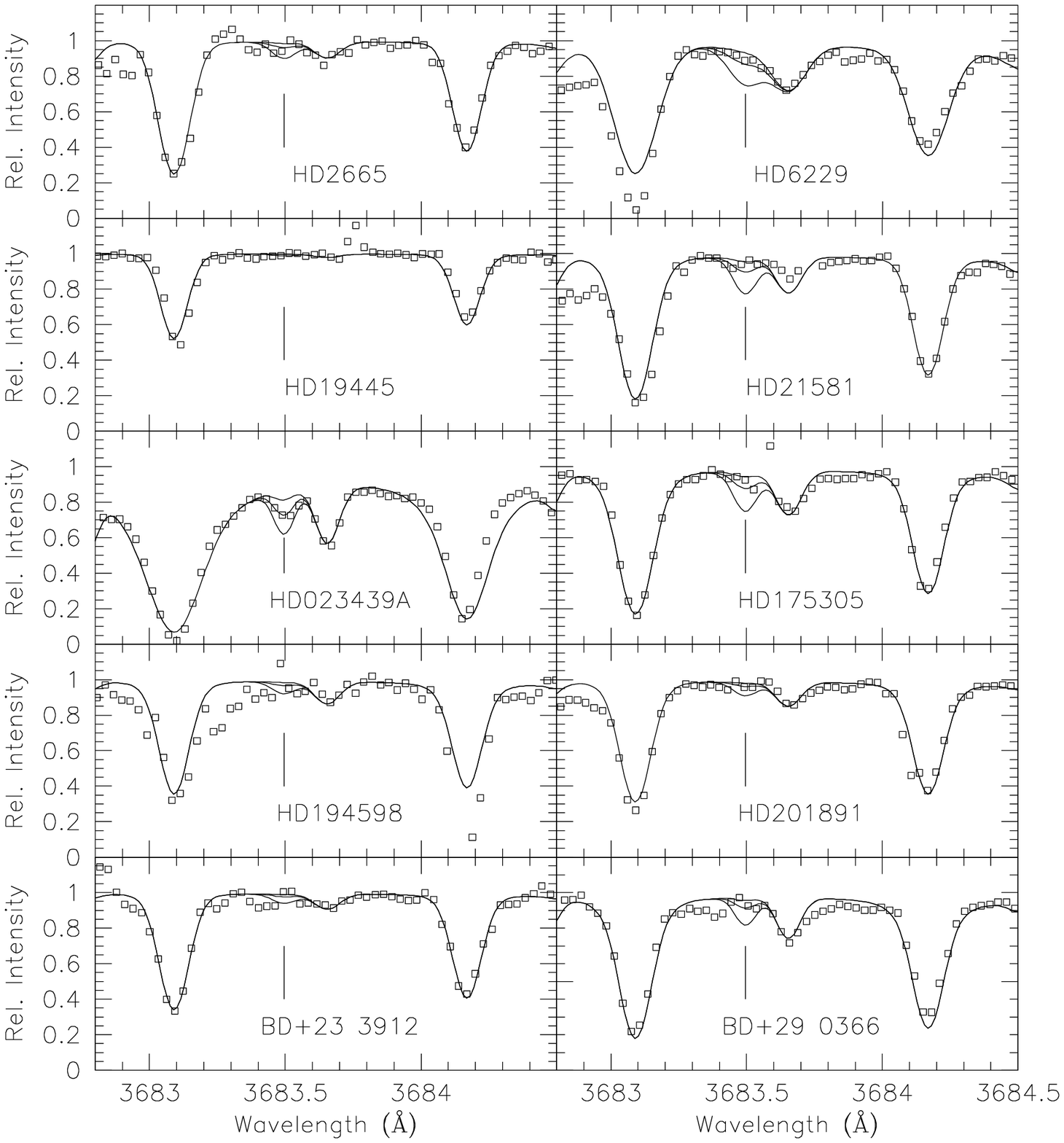}

\figcaption[fig3gratton.ps]{Comparisons between observed spectra for ten
metal-poor stars ({\it open squares}) and synthesized spectra for the same
stars ({\it solid lines}) for a small spectral region including the Pb~I
line at 3683.48~\AA. Synthetic spectra were computed with the atmospheric
parameters listed by Gratton et al.~(2000), and three different abundances
of Pb: [Pb/Fe]=0.0, 0.5, and 1.0. A mark signs the location of the Pb~I
line.\label{fig3}}

\newpage

\plotone{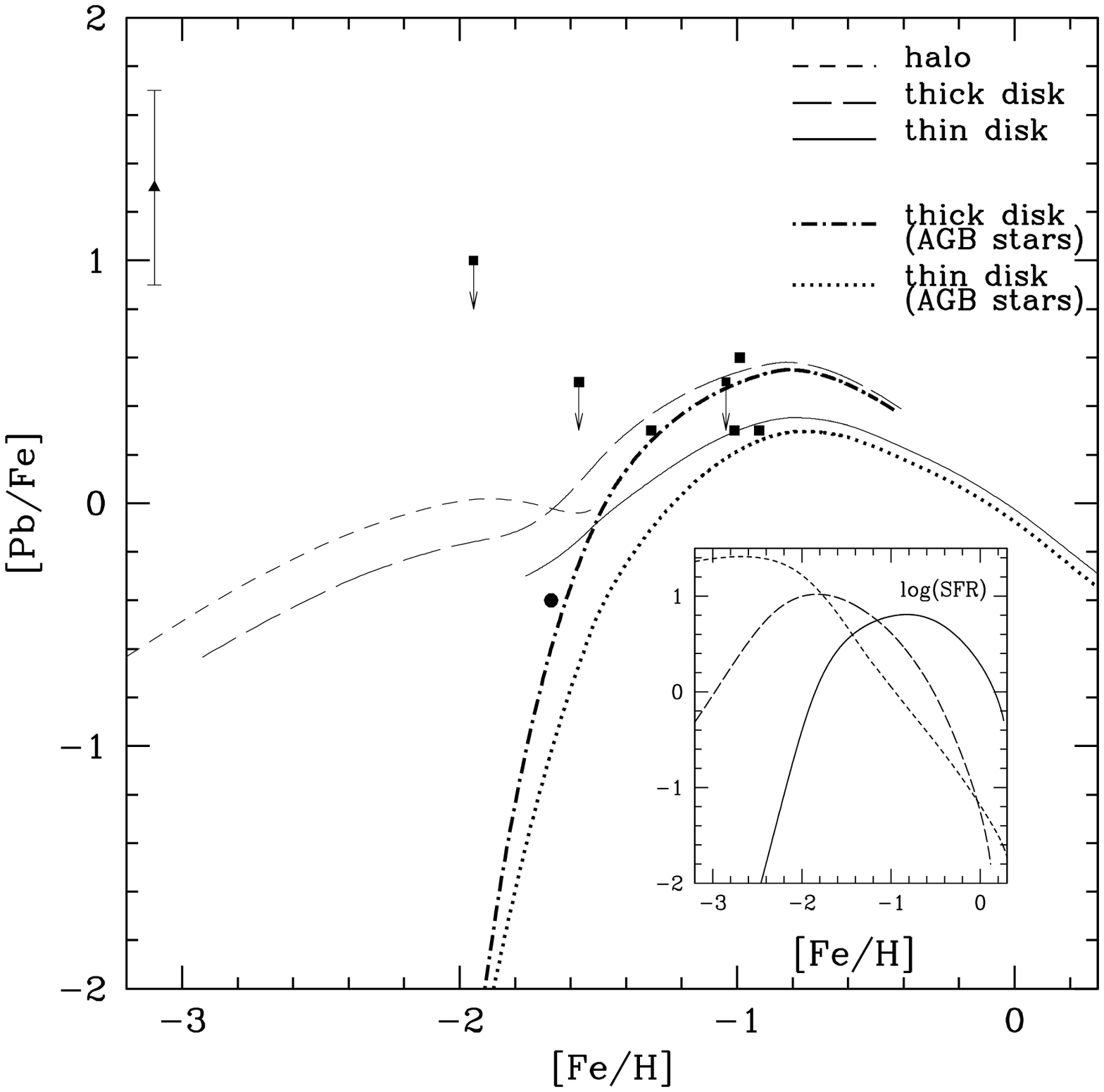}

\figcaption[fig4.ps]{Galactic evolution of [Pb/Fe] according to our
Galactic model predictions, including both $s-$ and $r-$process
contribution for the halo ({\it short-dashed line}), thick disk ({\it
long-dashed line}), and thin disk ({\it solid line}). The Pb $s$-fraction
from AGB stars in the thin disk is also shown with {\it thick-dotted
line}. Observational data are from Sneden et al.~(1998) ({\it filled
circle}), Sneden et al.~(2000) ({\it filled triangle}), and from the
present work ({\it filled squares}). Error bars are shown only when
reported for single objects.
In the {\it lower-right} small box is also shown Star Formation Rate
(in unit of \msb pc$^{-2}$ Gyr$^{-1}$ and logarithmic scale) for halo,
thick disk, and thin disk, according to our model predictions, as a
function of [Fe/H].\label{fig4}}

\end{document}